\documentclass{sigchi}

\CopyrightYear{2020}

\usepackage{balance}       
\usepackage{graphics}      
\usepackage[T1]{fontenc}   
\usepackage{txfonts}
\usepackage{mathptmx}
\usepackage[pdflang={en-US},pdftex]{hyperref}
\usepackage{color}
\usepackage{booktabs}
\usepackage{textcomp}
\usepackage{soul}
\usepackage{todonotes}

\usepackage{microtype}        
\usepackage{ccicons}          

\usepackage{todonotes}

\def\plaintitle{Towards a Practical Virtual Office for Mobile Knowledge Workers}
\def\plainauthor{Eyal Ofek, Jens Grubert, Per Ola Kristensen, Michel Pahud}

\def\plainkeywords{Virtual reality; virtual knowledge work; virtual office work; mobile virtual reality.}

\makeatletter
\def\url@leostyle{%
  \@ifundefined{selectfont}{
    \def\UrlFont{\sf}
  }{
    \def\UrlFont{\small\bf\ttfamily}
  }}
\makeatother
\def\pprw{8.5in}
\def\pprh{11in}

\definecolor{linkColor}{RGB}{6,125,233}
\hypersetup{%
  pdftitle={\plaintitle},
  pdfauthor={\plainauthor},
  pdfkeywords={\plainkeywords},
  pdfdisplaydoctitle=true, 
  bookmarksnumbered,
  pdfstartview={FitH},
  colorlinks,
  citecolor=black,
  filecolor=black,
  linkcolor=black,
  urlcolor=linkColor,
  breaklinks=true,
  hypertexnames=false
}

\begin{document}

\title{\plaintitle}

\numberofauthors{5}
\author{
	\alignauthor Eyal Ofek \\
	 \affaddr{Microsoft Research} \\
	\alignauthor Jens Grubert  \\
	 \affaddr{Coburg University of Applied Sciences} \\
	\alignauthor Michel Pahud    \\
		 \affaddr{Microsoft Research}\\
	\alignauthor Mark Phillips   \\
		\affaddr{Coburg University of Applied Sciences}\\
	\alignauthor Per Ola Kristensson  \\  
		\affaddr{University of Cambridge}\\
}

%
%
%
%
%

\teaser{
\centering 
  \includegraphics[width=1.9\columnwidth]{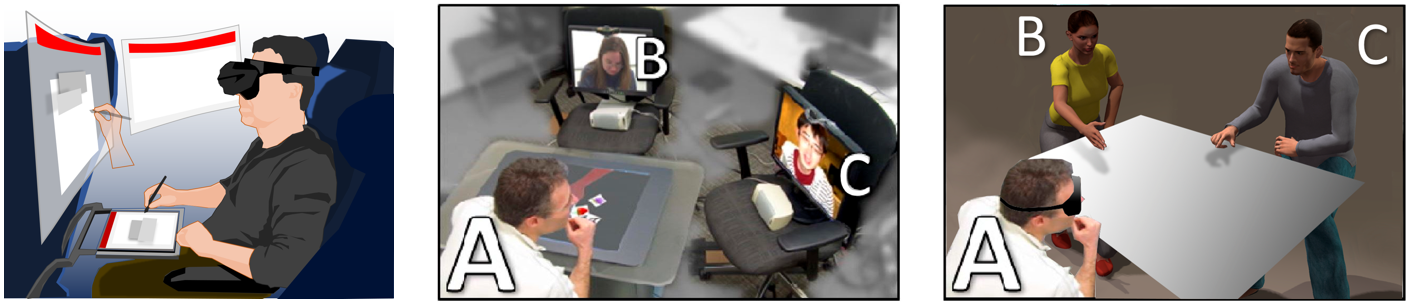}
  \caption{Left: Virtual Reality (VR) enables customized office environments, including large displays and background disturbance reduction, even in challenging environments such as airplanes. To this end, the interaction space is designed to be as small as possible to avoid interfering with the physical environment. Middle: A user interacts with remote participants in a physical office over a tabletop using physical screens as surrogates [59].  
  	Right: The same setting in VR with virtual representation of users.}
  \label{fig:teaser}
}

\maketitle

\begin{abstract}
As more people work from home or during travel, new opportunities and challenges arise around mobile office work. On one hand, people may work at flexible hours, independent of traffic limitations, but on the other hand, they may need to work at makeshift spaces, with less than optimal working conditions and decoupled from co-workers. Virtual Reality (VR) has the potential to change the way information workers work: it enables personal bespoke working environments even on the go and allows new collaboration approaches that can help mitigate effects of physical distance. In this paper, we investigate opportunities and challenges for realizing a mobile VR offices environments and discuss implications from recent findings of mixing standard off-the-shelf equipment, such as tablets, laptops or desktops, with VR to enable effective, efficient, ergonomic and rewarding mobile knowledge work. Further, we investigate the role of conceptual and physical spaces in a mobile VR office.
\end{abstract}









\section{Introduction}

The concept of the office as we know it is changing and this notion has only accelerated with the COVID-19 pandemic and the resulting wide-scale lock-downs that has suddenly forced millions of workers to carry out their work tasks in home environments that may be ill-suited for prolonged office work. The COVID-19 crisis has also highlighted the challenges in supporting effective collaborative environments and given rise to new problems, such as as ``zoom fatigue''\footnote{\url{https://www.nationalgeographic.co.uk/science-and-technology/2020/04/zoom-fatigue-is-taxing-the-brain-heres-why-that-happens} Last access July 2nd, 2020}, the phenomenon that extended videoconferencing tends to exhaust workers.

In general, more people are working in locations far from physical office building or while on the move. While this transition can induce many positive effects, such as  enabling flexible hours, reducing time spent in traffic and enable workers to live in far away locations, it does require workers to be able to carry out productive work in environments that might be less than optimal. The worker may need to use a small home office, or a makeshift work-space on the go, crumbed in an airplane seat, surrounded by a crowd of people with no privacy, or using a small desk in a hotel room. Furthermore, working from afar may strain collaboration with co-workers.

Another major change is the plurality of devices that users employ today. While in the past, most of the information work was done on stationary desktop PCs or relative mobile laptops that enabled a few hours of work but using limited input and a fixed small screen. Nowadays, user may carry ultra-mobile phones and tablets that can be used in small spaces but are limited in their input space and display sizes, as well as larger devices whenever there is more space or accessibility to power and WiFi. Cloud tools, enable to easily transfer data and documents between the devices, yet editing capabilities varies drastically between devices.

In recent years, VR technology has been progressed by leaps and bounds. Head-mounted devices (HMDs) have become light, cheap, supporting high-resolution displays that are on par with available screens (such as HP Reverb's 2160 x 2160 display), they may use inside-out optical tracking, which requires no special setups of the user's environment, enable optical hand tracking for controller-less interaction, support video pass-through for occasional interaction with the external world and be driven by existing laptops and tablets. While much of the hype around VR has focused on immersive gaming and entertainment, in this work, we focus on the use of mobile VR headset as a solution for many of the problems raised before. 

Past works such as the office of the future by Raskar et al. \cite{raskar1998office}, looked how to support immersive and fluid collaborative knowledge work. They utilized projection systems that increased the display area seen by the user and enabled the transition between different work stations. Building on this idea, without the need of instrumenting the user's environment, research has proposed to user consumer-oriented HMDs to extend the display space and mold a unified work space using multitude common devices, such as laptops and tablets \cite{grubert2018office}. We suggest a novel immersive work-space that enables the user to work in a similar fashion at a large variance of real physical environments. We direct our design space at the use of large display space but at the same time, small physical space that limits the input to the physical devices used by the user such as a tablet and their immediate vicinity. 

Furthermore, the sensing of VR HMDs enables extended collaboration than available by the devices alone. Beside sharing documents between co-workers and enabling co-edit of them, it is possible to render the reference space between multiple users for better collaborations \cite{tang2010three}, that is unlimited by physical limitations exist in real environments.

\section{Related Work}

There has been different works that relate to the mobile VR office work presented in this paper, mostly in the areas of mixed reality (MR), information
windows in spatial environments; and spatial interaction.

Early work on supporting knowledge workers using MR investigated the instrumentation of the office by projection systems to extend the interaction space in the office, e.g. \cite{kobayashi1998enhanceddesk} \cite{pinhanez2001everywhere} \cite{wellner1994interacting}. Later, research begun exploring MR for similar uses but focusing on the use of HMDs \cite{grubert2018office, guo2019mixed, ruvimova2020transport}. Different tasks, such as text entry (e.g., \cite{grubert2018text}, system control \cite{zielasko2019passive, zielasko2019menus} and visual analytics \cite{wagner2018virtualdesk} have been the focus of research. Those works were mostly looking at people working on a desktop PC, using the large interaction space of VR \cite{buschel2018interaction} with controllers or hand gestures \cite{kry2008handnavigator}, which do not fit the requirements for small interaction space of uncontrolled physical environment. 

The large display space of VR attracted researchers looking at organizing information around the user \cite{jetter2020vr, mcgill2020expanding}. From head and world references windows \cite{feiner1993windows}, arranging displays in a cylinder around the user \cite{billinghurst1999wearable}, to different coordinates system referring to the user's environment, and object, user's body or head \cite{laviola20173d}. Past works looked at fast access to virtual items \cite{chen2012extending, li2009virtual}, multitasking \cite{ens2014personal} and visual analytics \cite{ens2016spatial}. 

While VR enables large display space in any physical environment, most past works did not addressed the limitation of small input space that may be available for the user. We are looking at different ways in which we may increased the expressiveness of the user without using large gestures. While current devices such as tablets and phones enable only 2D display and touch input, we are looking at extending the information display and interaction in the depth direction too, 
 displaying different layers of information, and use the interaction space only or near the tablet in order to support interaction in constrained physical spaces \cite{Biener2020Breaking, Gesslein2020Pen, grubert2018office, mcgill2015dose}.

We draw on these rich sources of interaction ideas \cite{laviola20173d, argelaguet2013survey, mendes2019survey, grubert2016challenges, brudy2019cross} and adopt techniques in the context of VR interaction with touchscreens for mobile
knowledge workers. Our work complements multimodal techniques
combining touch and mid-air \cite{hilliges2009interactions, muller2014mirrortouch}, gaze-based techniques \cite{pfeuffer2017gaze+, hirzle2019design}
and ideas for combining HMDs with touchscreens \cite{grasset2007mixed, normand2018enlarging}
through novel techniques for accessing virtual windows around or
behind a physical touchscreen.

\section{Design Aspects for Mobile VR Work}
VR headsets can decouple workers from their physical environments and transport them the virtual environment, thereby allowing the user to work in a private virtual office controlled by them. The system may map limited input space to full control of the large display space (see Figure \ref{fig:teaser} left).
It may enable collaboration between far away workers in a way that is not possible in a physical space. In the following subsections we will describe these advantages.



\subsection{Work Environment Control}

As described by Grubert et al. \cite{grubert2018office}, the available work environment for the worker on the go may be sub-optimal, including obstacles for interaction in the vicinity of the user, lack of good illumination, a multitude visual and audio disturbances, lack of privacy and more. Using VR HMDs, users can generate their familiar ideal environments of their liking reducing to adapt to new context, layout around them, masking them for external visual and audio disturbances. While many virtual application use a large input space and enable people to roam inside a virtual space and operate using the full reach of their arms, in this paper we are focusing on using small input space and in particular use input spaces of devices such as a tablet, due to the following reasons: The use of large input gestures comes at the price of the amount of energy invested by the users \cite{hincapie2014consumed}, and there are situations such as in an airplane or touchdown spaces  where there is not enough room to do ample movements. To enable the user to work a full work day and reduce the exhaustion, we expect the user to do small gestures using supported hands, very similar to the gestures she would have done near her desk at the office. In fact, in the virtual world, which may not adhere to the physical laws of reality, small user gestures may reduce the amount of work users are needed to do today, such as locomotion, reaching toward far objects. Also, while the virtual display space maybe as large as we wish, interaction space may be limited by the physical environment. By designing interaction for a small input space, the user may be able to keep his familiar working gestures and muscle memory in many different physical environments. 

\subsection{Privacy and Social Acceptability}
\begin{figure}
  \centering 
  \includegraphics[width=0.9\columnwidth]{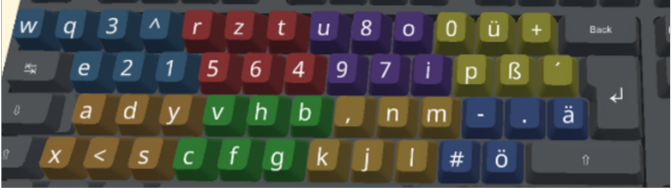}
  \caption{Visually Scrambling the layout of a virtual keyboard, enables the user to type a password without worry of other people in his physical vicinity [55].}
  \label{fig:Scrable}
\end{figure}

One critical design aspect for VR office workers situated in uncontrolled environments, such as shared office spaces, trains, airplanes, cafes or public spaces, is preventing eavesdropping of at least highly sensitive information, such as passwords. While VR HMDs are inherently personal and do not expose the displays to people around the user, the input gestures of the user are exposed to the public around her. Schneider et al. \cite{schneider2019reconviguration} addressed the use of standard keyboards while working in VR. By augmenting the arrangement of the keyboards as seen in the VR space they obfuscate the entry of passwords by the user \ref{fig:Scrable}. In a similar fashion, the use of private display space that can be rearranged without the knowledge of an external viewer to hide the semantics of inputs such as stylus strokes or soft keyboards to protect sensitive data input.  

When using VR HMDs in public spaces, one also needs to consider the social acceptability of those interactions \cite{gugenheimer2019challenges}. For example, Schwind et al. \cite{schwind2018virtual} indicated, that VR HMDs are accepted to be used in public spaces, but only when social interaction between physical present people is not expected. For example, users acceptance of VR HMDs in settings such as public transportation (trains or busses) was significantly higher than in a public cafe. Williamson et al. \cite{williamson2019planevr}, highlighted the need for allowing for transitioning between physical and virtual worlds to allow to accommodate dealing with interruptions when using VR HMDs on airplanes. Eghbali et al. \cite{eghbali2019social}, put forward design recommendations for socially acceptable use of VR in public spaces, again recommending to be able to interact with the  physical space and users in VR and ideas how to achieve this have been proposed \cite{mcgill2015dose, gugenheimer2017sharevr, hartmann2019realitycheck}.

\subsection{Re-appropriating Input Devices}
\begin{figure}
  \centering 
  \includegraphics[width=0.9\columnwidth]{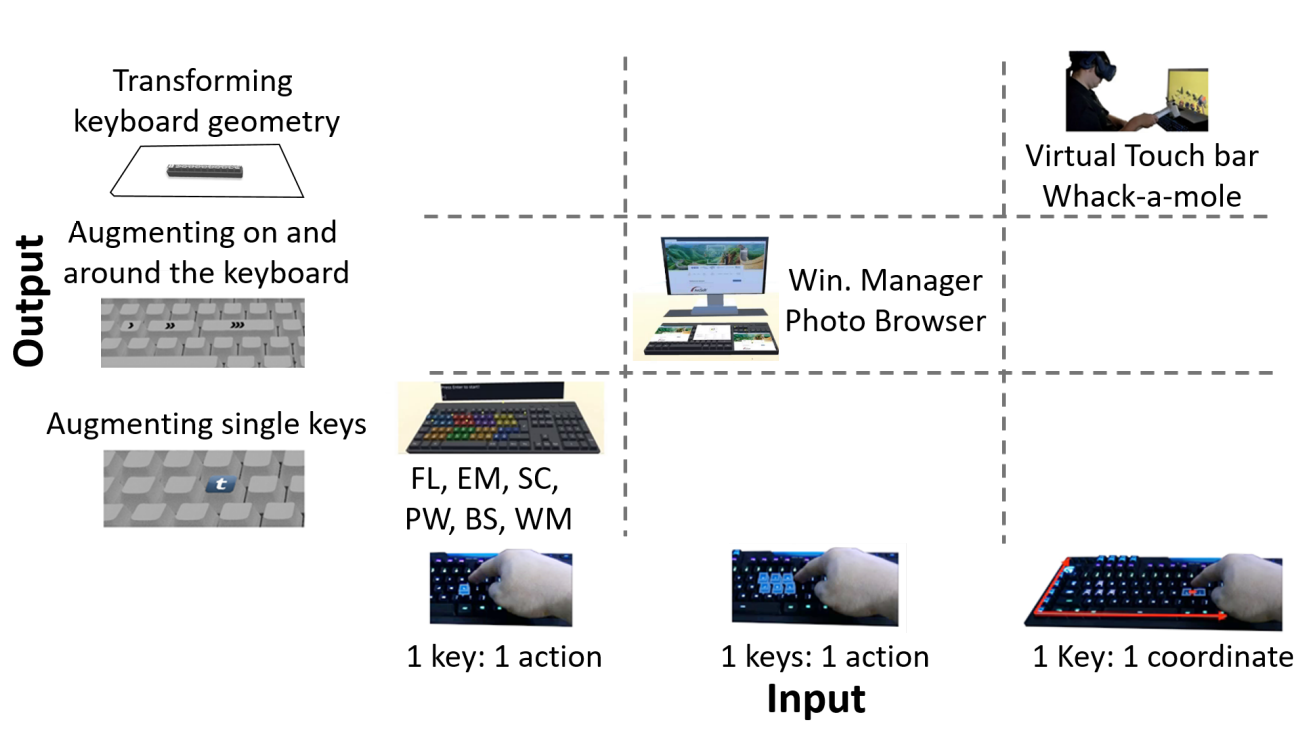}
  \caption{Input-output dimensions of reconfiguring physical keyboards in virtual reality with mapped example applications. The x-axis shows input mappings and the y-axis shows output mappings [55].} 
  \label{fig:InOut}
\end{figure}

Information worker uses a slew of input devices, some of them have little changed over tens of years. For example, the physical keyboard is still the most known and preferred input device for text entry. The use of a VR HMD obscure the view of the keyboard from the user, and the application renders a virtual view of the keyboard. Since the virtual environment is not bound the limitation of the physical world, it is possible to change the look and functionally of the device according to application needs \cite{schneider2019reconviguration}. Figure \ref{fig:InOut} shows two different dimensions in which a physical keyboard may be augmented for different use in VR. The output mapping, or the way the keyboard is rendered in VR can change the functionalities of keys on the keyboard. Some keys may not be displayed to direct the user to specific keys, some keys may be rendered as one big key for easier selection, or just changing the functionalities of keys. Another option is to use a small portable keyboard, and change it's keys functionality according to need to simulate a full size keyboard. The input dimension changes the understanding of a signal from the keyboard. While the default semantics is that each key has a unique input signal, it is also possible to combine several keys to one big key or even use the full keyboard to sense 2D location of a press.

The same type of augmentation may be used on other devices, whether it is a mouse or track pad a touch screen or a stylus. for example, adding new semantics for mouse buttons, wheel or movement directions, or a stylus that represents it's functionality. 

\subsection{Retargeting Visuals of Input Space}

\begin{figure}
  \centering 
  \includegraphics[width=0.9\columnwidth]{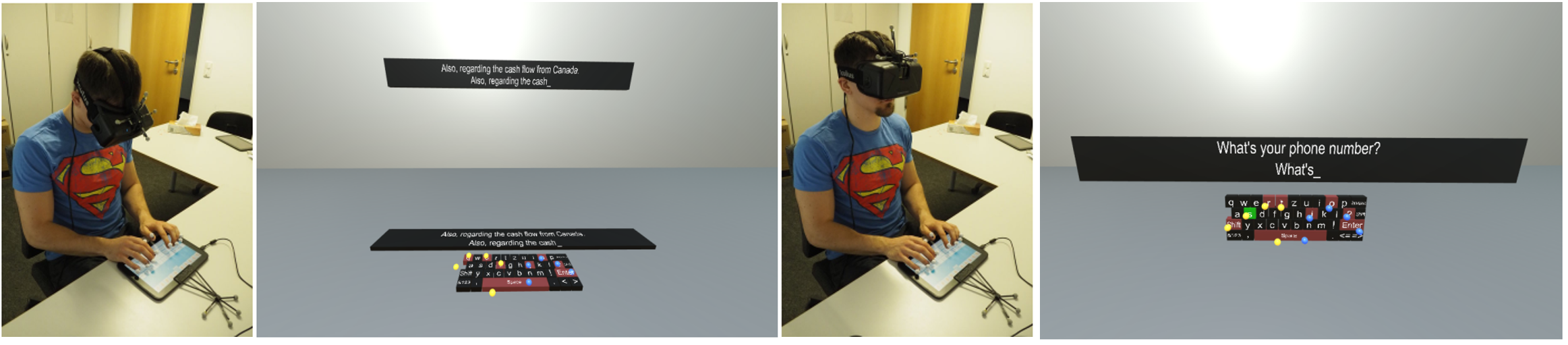}
  \caption{Retargeting the input space and the user hand to a location easier seen by the user [21].}
  \label{fig:Retarget}
\end{figure}

The limited input space may not allow for comfortable viewing. For example, when interacting with a tablet laying on a tray of an airplane seat, it may not be easy to tilt the HMD down to look at the tablet and the hands around it. Following the work of Grubert et al.~\cite{grubert2018text}, the display of the tablet and the user hands may be retargeted from their physical space to lie in the user's field of view, see Figure \ref{fig:Retarget}. Grubert et al~\cite{grubert2018text} found that if the user is typing on a physical keyboard there is no performance penalty in such retargeting, and when using a soft keyboard that lacks of haptic feedback, there is a minor speed penalty, but it enables much improved ergonomics for the user's head.

Such a difference between the input space and it's corresponding display may enable better utilization of the physical affordances of the user's environment, regardless of the display space. For example, changing a continuous parameter, such as audio volume, can be achieved by moving a slider on a tablet. However, this requires the user to target their finger to the slider and visually determine the new position for the slider. The extended sensing of the HMDs may allow for tracking users' fingers while they are away from the tablet, and re-appropriate natural physical landmarks, such as a handle of a chair or the side edge of a tablet or a tray, to guide the finger as it moves the slider.

\subsection{Sharing Reference Spaces}

When people collaborate in physical environments, we can define different spaces of collaboration \cite{tang2010three}: {\em Task space} is where the work appears. For the information worker this is usually a set of documents such as spreadsheets, text documents, shared whiteboard, etc.. Most modern applications let multiple people edit the same document concurrently online. The user may see the locations of the other people edits, send comments between users and so on. {\em Person space} are where verbal and facial cues are used for expression. Video and audio chat applications can be used to realize this level of communication. However, in physical environment, people also use {\em reference space} where remote parties can use body language, such as pointing, to refer to the work, organize the separation of work to different participants and more. Figure \ref{fig:teaser}, middle, shows an office that has these three spaces with a local participant A and remote participants B and C. Figure \ref{fig:teaser} right, shows the same office virtually created when user A wears a HMD.

The use of VR HMDs opens up several opportunities for communication of the reference space, some even beyond the collaboration in a physical environment. 
For example, the pointing direction of users hands or gaze can be visualized for highlighting relevant objects and facilitating collaboraiton in VR \cite{sato2001visualizing, pfeiffer2008conversational, piumsomboon2019effects, wang2019head}. 
Tang et al. \cite{tang2010three} looked at different arrangements for collaborators around a table {\em Task space}. They found that if people sitting around the document as they tend to do in a physical environment are visualized, it is easy to associate the gestures with people, however each person sees the document from a different direction. They also tried to render all users hands as if they all sit in one locations, but associating the hands with different users proved to be confusing. VR HMDs enable new arrangements, where all collaborators may see the task space from the same point of view, but translating their avatars and hand gestures, as if they are sitting around the task space for easy associations. 

Further, as people tend to work from a far, there is a starting option of virtual room mates: People that are rendered as working in the user's vicinity, and if the user is moving her body toward a virtual neighbour it can originate conversations and ad-hoc collaborations. If on the other hand the user is concentrating on a specific problem, it may fade out the neighbours to help the user concentration.

\subsection{Architectural Space}
While VR has a long tradition of being used as design tool for constructing architectural spaces \cite{wang2008mixed}, designing purely virtual worlds for productive use from an architectural perspective has not seen detailed attention, yet. Often, virtual office environments have been modelled to be replicas from physical rooms \cite{knierim2018physical}, or abstract office spaces \cite{grubert2018text,grubert2018effects}. Only recently, researchers have begun to investigate the effects of virtual space and place on office work. For example, Ruvimova et al.~\cite{ruvimova2020transport} indicated that VR environments (like working on a virtual beach) can outperform work done in a physical open office in terms of flow and performance. Guo et al.~\cite{guo2019mixed} were inspired by insights on the positive effects of elements of nature on workers well-being~\cite{an2016we} and used a naturalistic environment for studying long term work in a VR office. Still, a number of research questions remain to fully utilize the power of well designed VR spaces for mobile knowledge workers such as:

\textit{How much congruence between physical and virtual space do office workers need to work productively, safely, healthy and enjoyably?} When working in a physical environment, co-located with other users, questions about awareness in VR office environments rise \cite{mcgill2015dose, huang2018improving, kanamori2018obstacle}. Also, transitions between physical and virtual world have been studied , e.g., in terms of presence \cite{kijima1997transition, steinicke2009does, steinicke2010gradual, valkov2017smooth, jung2018limbo}. However, the effects of awareness and transition techniques on flow and productivity of knowledge workers have just begun to be explored \cite{ghosh2018notifivr, rzayev2019notification}. Also, it remains an open challenge, how spatial design can support communication both with physically co-located and remote users.

\textit{What is the impact of architectural design on virtual office environments?} Large IT companies have invested substantial resources to design and build physical office buildings such as Amazon spheres, Facebook Menlo Park, Apple Park or Google Mountain View. Aspects such as spatial dimensions, visibility, materials and lighting or the soundscape of an environment have seen great attention to facilitate chance encounters and physical meetings and still allow workers to focus on individual tasks, if needed. This has partly been done due to the observations that architectural design can support collaboration, productivity and creativity  \cite{brill1984using, duffy2005impact, waber2014workspaces, sailer2019correspondence}. However, with recent trends to increase work from the home office, even with its positive effects on workers \cite{hill2003does}, these goals are at stake. Groupware tools such as Slack, Microsoft Teams or Workplace by Facebook focus at productivity and hardly replace planed or chance personal interactions. It remains an open question if properly spatially design virtual office environments can have similar effects on collaboration as physical office environments. To this end, studies are needed which carefully compare effects of architectural and virtual space. This is especially challenging, as potential confounds such as technology artefacts \cite{bowman2007virtual} or user embodiment \cite{latoschik2017effect} have to be considered as well.

Another open question to study is related to how a virtual environment could create opportunities and motivate participants to have unscheduled casual meetings such as hallway conversations, which can be a great contribution to a project, reinforce the bounds between colleagues, and decrease the isolation factor for remote workers. Hallway conversations could also encompass side discussions between team members next to the conference room that happen just after a meeting. Experiences such as having a virtual hallway nearby the virtual office could be interesting to study.



\section{Conclusions}
As an increasing number of people work away from the offer there is an opportunity to use VR to enhance they way we work and collaborate. VR can, for example, disconnect the worker from disturbances, allow the office worker to virtually recreate a familiar working environment, and enable new collaborative ways of working to improve efficiency and mitigate the effect of people working at a distance. In this paper, we have reported on recent findings in this emerging area and discussed design aspects that need to be taken into account to realize a practical mobile virtual office: work environment control, privacy and social acceptability, re-appropriation of input devices, retargeting visuals of input space, sharing reference spaces and architectural space. Through this we hope to spark further research in this rapidly expanding field.

\balance
\bibliographystyle{SIGCHI-Reference-Format}
\bibliography{vrmobilework}

\end{document}